\documentclass[a4paper,10pt]{article}
\usepackage[T1]{fontenc}
\usepackage[latin9]{inputenc}
\usepackage{color}
\usepackage{float}
\usepackage{amsmath}
\usepackage{graphicx}
\usepackage{cite}

\title{MAXIMIZING POSITIVE OPINION INFLUENCE USING AN EVIDENTIAL APPROACH}
\author{Siwar Jendoubi$^{1,2,3}$, Arnaud Martin$^2$, Ludovic Li\'etard$^2$,\\ Hend Ben Hadji$^3$ and Boutheina Ben Yaghlane$^4$\\
\\
$^1$Universit\'e de Tunis, ISG Tunis, LARODEC, \\E-mail: jendoubi.siwar@yahoo.fr\\
$^2$Universit\'e de Rennes 1, IUT Lannion, IRISA, \\ E-mail: \{Arnaud.Martin,Ludovic.lietard\}@univ-rennes1.fr\\
$^3$ Centre d'Etudes et de Recherche des T\'el\'ecommunications, \\E-mail: hend.benhji@cert.mincom.tn\\
$^4$Universit\'e de Carthage, IHEC Carthage, LARODEC, \\E-mail: boutheina.yaghlane@ihec.rnu.tn
}

\begin{document}
\date{}

\maketitle

\begin{abstract}
In this paper\footnote{These research works and innovation are carried out within the framework of the device MOBIDOC financed by the European Union under the PASRI program and administrated by the ANPR. Also, we thank the ``Centre d'Etude et de Recherche des T{\'e}l{\'e}communications'' (CERT) for their support.}, we propose a new data based model for influence maximization
in online social networks. We use the theory of belief functions to overcome the data imperfection problem. Besides, the proposed model searches to detect influencer users that adopt a positive opinion about the product, the idea, etc, to be propagated. Moreover, we present some experiments to show the performance of our model.
\end{abstract}

\section{Introduction}

The phenomena of influence propagation through social networks have attracted a great body of research works. A key function of an online social network (OSN), besides sharing, is that it enables users to express their personal opinions about a product or trend of news by means of posts, likes/dislikes, etc. Such opinions are propagated to other users and might make a significant influence on them, either positive or negative. 
 
Real world is full of imprecision and uncertainty and this fact necessarily impacts on OSN data. In fact, social interactions can not be always precise and certain, also, OSN allows only a limited access for their data which generates more imprecision and uncertainty. Then, if we ignore this imperfection, we may  be confronted to erroneous analysis results. In such a situation, the theory of belief functions \cite{Dempster67a,Shafer76} have been widely applied. Furthermore, this theory was used for analyzing social networks \cite{Jendoubi14a,Jendoubi2015,Zhou2015}.

Influence maximization (IM) is the problem of finding a set of $k$ seed nodes that are able to influence the maximum number of nodes in the social network. In the literature, we find many solutions for the IM problem. \textit{Kempe et al.} \cite{Kempe03} propose two propagation simulation models which are the \textit{Linear Threshold Model (LTM)} and the \textit{Independent Cascade Model (ICM)}. Besides, the credit distribution (CD) model \cite{Goyal12} is a data based approach that investigates past propagation to detect influencers. However, these solutions does not consider the user's opinion. Zhang et al. \cite{Zhang2013} propose an opinion based cascading model that considers the user's opinion about the product. However, their work is not based on real word data to estimate user's opinion and influence.  

In this paper, we propose a new data based model for influence maximization
in online social networks that searches to detect influencer users
that adopt a positive opinion about the product. 
The proposed model is data based because we use past propagation to estimate the influence, and users messages to estimate the opinion. Besides, it uses the theory of belief functions to estimate the influence to deal with
data imprecision and uncertainty. To the best of our knowledge, the proposed model
is the first evidential data based model that maximizes the influence
on OSN, detects influencer users having a positive opinion about
the product and uses the theory of belief functions to process the
data imperfection.

The remainder of this paper is organized as follows: section 2 introduces the proposed model
for maximizing the positive opinion influence, section 3
shows the performance of our model through some relevant
experiments. Finally, the paper is concluded in section 4.

\vspace{-0.05cm}

\section{Maximizing positive opinion influence}

In this section, we present our positive opinion influence measure and the proposed influence maximization
algorithm.

\subsection{Influence measure}

Given a social network $G=\left(V,E\right)$, a frame of discernment expressing opinion $\Theta=\left\{ Pos,\, Neg,\, Obj\right\} $, $Pos$ for positive, $Neg$ for negative and $Obj$ for objective, a frame of discernment expressing influence and passivity $\Omega=\left\{ I,P\right\} $, $I$ for influencer and $P$ for passive user, a probability distribution $\Pr^{\Theta}\left(u\right)$ defined on $\Theta$ that expresses the opinion of the user $u\in V$ about the product and a basic belief assignment (BBA) function~\cite{Shafer76}, $m^{\Omega}\left(u,v\right)$, defined on $\Omega$ that expresses the influence that exerts the user $u$ on the user $v$. The first step of the influence maximization process is to measure the influence of each user in the network. Then we propose an influence measure to estimate the positive influence of each user in the network. 

The mass value $m^{\Omega}\left(u,v\right)\left(I\right)$ measures the influence of $u$ on $v$ but without considering the opinion of $u$ about the product. We define the positive opinion influence of $u$ on $v$ as the positive proportion of $m^{\Omega}\left(u,v\right)\left(I\right)$ and we measure this proportion as: 
\begin{equation}
Inf_{Pos}\left(u,v\right)=\textrm{Pr}^{\Theta}\left(u\right)\left(Pos\right).m^{\Omega}\left(u,v\right)\left(I\right)
\end{equation}

Next, we define the amount of influence given to a set of nodes $S\subseteq V$
for influencing a user $v\in V$. We estimate the influence of $S$
on a user $v$ as follows:

\begin{equation}
Inf_{Pos}\left(S,v\right)=\begin{cases}
1 & \textrm{if} \, v\in S \\
{\displaystyle \sum_{u\in S}{\displaystyle }\sum_{x\in D_{IN}\left(v\right)\cup\left\{ v\right\} }Inf_{Pos}\left(u,x\right).Inf_{Pos}\left(x,v\right)} & \textrm{Otherwise}
\end{cases}
\end{equation}
such that $Inf_{Pos}\left(v,v\right)=1$ and $D_{IN}\left(v\right)$ is
the set of nodes in the indegree of $v$. Finally, we define the influence
spread $\sigma\left(S\right)$ under the evidential model as the total
influence given to $S\subseteq V$ from all nodes in the social network
as $\sigma\left(S\right)=\sum_{v\in V}Inf_{Pos}\left(S,v\right)$. In the
spirit of the IM problem, as defined by \textit{Kempe et al.} \cite{Kempe03}, $\sigma\left(S\right)$
is the objective function to be maximized.

\subsection{Influence maximization}

In this section, we present the evidential positive opinion influence
maximization model. Its purpose is to find a set of nodes $S$ that
maximizes the objective function $\sigma\left(S\right)$. Given a
directed social network $G=\left(V,\, E\right)$, an integer $k\leq |V|$, the goal is to find a set of users $S\subseteq V,$ $|S|=k$,
that maximizes $\sigma\left(S\right)$. We proved that $\sigma\left(S\right)$
is monotone and sub-modular, also the influence maximization under
the proposed model is NP-Hard. However, the number of pages limitation
prevents us to present proofs in detail. 

The influence maximization under the evidential positive opinion influence
maximization model is NP-Hard, consequently, the greedy algorithm performs
good approximation for the optimal solution especially when we use
it with this formula: 
\begin{equation}
\label{eq:mg}
\sigma_{Bel} \left(S\cup\left\{ x\right\} \right)-\sigma_{Bel}\left(S\right)=1+\sum_{v\in V\setminus S\,}\sum_{a\in D_{IN}\left(v\right)\cup\left\{ v\right\} }Inf\left(x,a\right).Inf\left(a,v\right)
\end{equation}
that computes the marginal gain of a candidate node $x$. We choose
the cost effective lazy-forward algorithm (CELF) \cite{Leskovec07b}
which is a two pass modified greedy algorithm. It exploits the sub-modularity
property of the objective function, also, it is about 700 times faster
then the basic greedy algorithm. 
The CELF based evidential influence maximization algorithm starts by estimating the marginal gain of all users in the network and sorts them according to their marginal gain, then, it selects the user that have the maximum marginal gain and add it to $S$ (seed set). After that, the algorithm iterates on the following steps until getting $|S| = k$: 1) Choose the next user in the list, 2) Update its marginal gain (formula (\ref{eq:mg})), and 3) If the chosen node keeps its position in the list (it still the maximum)
then add it to $S$

\section{Experiments}

In this section, we conduct some experiments on real world data. We used the library Twitter4j\footnote{http://twitter4j.org/en/index.html} which is a java implementation of the Twitter API to collect Twitter data. We crawled the Twitter network for the period between 08/09/2014 and 03/11/2014, and we filtered our data by keeping only tweets that talk about smartphones and users that have at least one tweet in the data base. To estimate the opinion polarity of each tweet in our data set, first, we used the java library ``Stanford POS Tagger'' \footnote{http://nlp.stanford.edu/software/tagger.shtml} with the model ``GATE Twitter part-of-speech tagger''\footnote{https://gate.ac.uk/wiki/twitter-postagger.html} that were designed for tweets. This step gives a tag (verb, noun, etc) to each word in the tweet. After, we estimated the opinion polarity of each tweet using the SentiWordNet 3.0\footnote{http://sentiwordnet.isti.cnr.it/} dictionary and tags from the first step. We estimated $m^{\Omega}\left(u,v\right)$ using the network structure and past propagation between $u$ and $v$. First, we calculated the number of common neighbors between $u$ and $v$, the number of tweets where $u$ mentions $v$ and the number of tweets where $v$ retweets from $u$. After we used the process defined by \textit{Wei et al.} \cite{Wei13} to estimate a BBA for each defined variable. Finally we combine the resulting BBAs to obtain $m^{\Omega}\left(u,v\right)$. In this section, we call belief model: our model in which we use $Inf\left(u,v\right)=m^{\Omega}\left(u,v\right)\left(I\right)$ as influence measure, CD model: the credit distribution model and opinion model: the proposed positive opinion based model.

\begin{figure}
\begin{centering}
\includegraphics[scale=0.45]{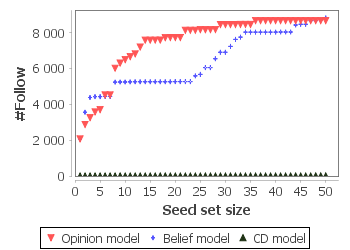} \includegraphics[scale=0.45]{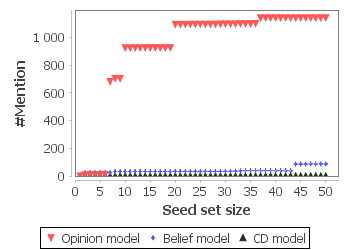}
\par\end{centering}

\begin{centering}
\includegraphics[scale=0.45]{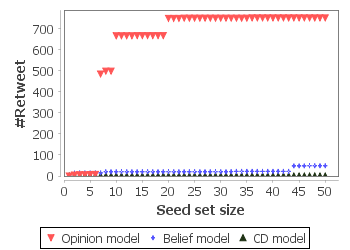} \includegraphics[scale=0.45]{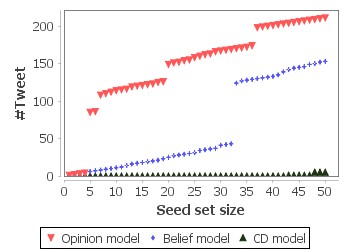}
\par\end{centering}

\caption{Comparison between opinion model, belief model and CD model according to \#Follow,
\#Mention, \#Retweet and \#Tweet\label{fig:Comparison}}
\end{figure}

The goal of the first experiment is to show that the proposed model
detects well influencer spreaders. To examine the quality of the selected
seeds, we fixed four comparison criteria which are: the number of
followers, \#Follow, the number of tweets, \#Tweet, the number of
times the user was mentioned and retweeted, \#Mention and \#Retweet.
In fact, we assume that if a user is an influencer on Twitter he would be necessarily:
very active so he has a lot of tweets, he is followed by many users
in the network, he is frequently mentioned and his tweets are retweeted several times. In Figure
(\ref{fig:Comparison}), we compare the maximization results of the proposed opinion model with CD model and belief model according to the fixed criteria. Figure (\ref{fig:Comparison}) shows the performance
of the proposed model against CD model and belief model. In fact, we see that
the proposed opinion model detects influencer that 
have many followers (more than 8000 for 50 influencer), many tweets (over 250 for 50 users), many mentions (about 1200) and many retweets (about 800).
However, users detected using the belief model have only two good criteria, \textit{i.e.} \#Follow (over 8000 follower for 50 users) and \#Tweet (over 150 tweets for 50 users), and the CD model does not satisfy any criteria. This shows that, the opinion model is the best in detecting influencers.

In a second experiment, we calculated the mean positive opinion of the first 100 influencer user. The proposed model performed well by selecting influencers that have a positive opinion about the product. In fact, it gives a mean positive opinion equals to 0.89 ($\pm0.04$, $95\%$ confidence interval). However, the belief model gives 0.34 ($\pm0.05$) and the CD model gives only 0.09 ($\pm0.04$). These results show the performance of the proposed model in selecting influencer users that have a positive opinion against the belief and the CD models that have not.

\section{Conclusion}

In this paper, we proposed a new influence measure that estimates the positive opinion
influence of OSN users. We used the theory of belief functions to deal with the problem of data imperfection. In future works, we will search to improve the proposed influence maximization model by considering other parameters like the user's profile and the propagation time.

\bibliographystyle{IEEEtran}
\bibliography{biblio}

\end{document}